# Title: Quantized chiral edge conduction on reconfigurable domain walls of a magnetic topological insulator

**Authors:** K. Yasuda[1*], M. Mogi[1], R. Yoshimi[2], A. Tsukazaki[3], K. S. Takahashi[2], M. Kawasaki[1,2], F. Kagawa[2], Y. Tokura[1,2*]

**Affiliations:**

[1]Department of Applied Physics and Quantum-Phase Electronics Center (QPEC), University of Tokyo, Tokyo 113-8656, Japan.

[2]RIKEN Center for Emergent Matter Science (CEMS), Wako 351-0198, Japan.

[3]Institute for Materials Research, Tohoku University, Sendai 980-8577, Japan.

*Correspondence to: yasuda@cmr.t.u-tokyo.ac.jp, tokura@riken.jp

**Abstract**: The electronic orders in magnetic and dielectric materials form the domains with different signs of order parameters. The control of configuration and motion of the domain walls (DWs) enables gigantic, nonvolatile responses against minute external fields, forming the bases of contemporary electronics. As an extension of the DW function concept, we realize the one-dimensional quantized conduction on the magnetic DWs of a topological insulator (TI). The DW of a magnetic TI is predicted to host the chiral edge state (CES) of dissipation-less nature when each magnetic domain is in the quantum anomalous Hall state. We design and fabricate the magnetic domains in a magnetic TI film with the tip of the magnetic force microscope, and clearly prove the existence of the chiral one-dimensional edge conduction along the prescribed DWs. The proof-of-concept devices based on the reconfigurable CES and Landauer-Büttiker formalism are exemplified for multiple-domain configurations with the well-defined DW channels.

**One Sentence Summary:** Magnetic-tip assisted domain writing and *in situ* transport measurements have unraveled the quantized chiral edge conduction on the reconfigurable DWs of a magnetic TI

**Main Text:** Topological phases of matter have been conceptually extended in the condensed matter science ever since the discovery of the quantum Hall effect (QHE) (*1*) The theoretical prediction and the experimental discovery of two- or three-dimensional topological insulators (TIs) triggered the emergence of wide varieties of topological quantum materials and topologically nontrivial phenomena (*2*,*3*). Topological phases are characterized by integer indices defined in the bulk electronic state, while gapless quasiparticle excitation appears at the boundary of topologically distinct phases either at the surface (edge) or interfaces, affording the bulk-edge correspondence. For example, in a two-dimensional electron system under magnetic field, the QHE is characterized by a filling factor of Landau levels, where corresponding numbers of chiral edge states (CESs) appear at the sample edge. Such CES is experimentally detected not only at the sample edge (*1*,*4*) but also at the boundary of different filling factors, *e.g.* in in-plane *p-n* junction of graphene formed with local gating under fixed magnetic field (*5*,*6*).

Topological phases are further enriched by the notion of spontaneous symmetry breaking. Beyond the prerequisites of a high magnetic field to induce the QHE, the spontaneous magnetization induces the CES at the sample edge as well as at the domain wall (DW) in the quantum anomalous Hall state under zero magnetic field (*2*,*3*,*7*). The quantum anomalous Hall effect (QAHE) has recently been found in three-dimensional TI $(Bi_{1-y}Sb_y)_2Te_3$ thin films doped with magnetic ions, Cr (*8-11*) or V (*12*), having perpendicular magnetic anisotropy. When the Fermi energy is tuned within the magnetization induced mass gap, the Hall conductance $\sigma_{xy}$ is quantized to $+e^2/h$ or $-e^2/h$, which evidences the occurrence of QAHE with the formation of the CES at the sample edges. Here, the QAHE is characterized by a topological number called a Chern number, which have a one-to-one correspondence to a conventional magnetic order parameter. Thus, the Chern number ($C = +1$ or $-1$) and the chirality of the CES can be controlled by the magnetization direction ($M > 0$ or $M < 0$). Furthermore, since the Chern number must discontinuously change at the DW between up and down magnetic domains, the CES is expected to appear also at the DW as schematically shown in Fig. 1, A and B (*2,3*). Notably, the position of the CES, in contrast to those at the sample edge (*1,4,8-12*) or at the local gate boundary (*5,6*), has a potential to be manipulated via domain control by external fields such as local magnetic field or current induced spin-orbit torque (*13,14*). Thus, a reconfigurable localized electronic channel based on the CES can be constructed with less dissipation and without application of magnetic field. In this study, superior to the previous approaches to probe the CES via naturally-formed uncontrollable DWs (*15,16*), the quantized conductance of the CES is clearly probed in a controllable way using magnetic force microscopy (MFM) technique.

Magnetic TI thin films, Cr-doped $(Bi_{1-y}Sb_y)_2Te_3$, are grown on InP substrates as shown in Fig. 1C, where Cr modulation doping stabilizes the QAHE up to about 1 K (*11*). Figure 1D shows the MFM image taken at 0.5 K with a non-contact mode (see Materials and Methods), visualizing naturally-formed multi-domain structure with up (red) and down (blue) magnetization and DWs (whitish region) in between. Bubble and stripe domain patterns are observed (fig. S1), typical for a thin film with perpendicular magnetic anisotropy (*17*). Here, we utilize the stray field from the MFM tip (fig. S2) to write the magnetic domain with an arbitrary shape as follows. We first set the respective magnetizations of the MFM tip and the sample in opposite directions (fig. S3). Then, we scan over the sample inside the dashed frame in Fig. 1E with a contact mode (see Materials and Methods) under a small magnetic field of 0.015 T (< the coercive field of the TI film $B_{c,TI}$). As can be seen in the MFM image taken with a non-contact mode (Fig. 1F), the magnetization direction only in the scanned area is reversed, meaning that the above procedure is effective for domain writing. By applying this writing technique to Hall-bar devices, the electrical detection of the CES is exemplified.

Figure 2A shows the magnetic field dependence of the transport property. In the single-domain states, the Hall resistance is $|R_{13}| = |R_{24}| = 25.1$ k$\Omega$ ($\sim h/e^2 = 25.8$ k$\Omega$) with low residual longitudinal resistance of $R_{12} = R_{34} = 1.8$ k$\Omega$ at $T = 0.5$ K, which represent the occurrence of QAHE. The sign of the Hall resistance is reversed at around $B_{c,TI} \sim 0.06$ T, reflecting the change of the magnetization direction and hence of the Chern number. In the multi-domain state around $B_{c,TI}$, the longitudinal resistance is continuously changed to take a peak structure. The symmetric resistance values $R_{13} = R_{24}$ and $R_{12} = R_{34}$ in the Hall-bar shape are typical behavior in QAHE (*8-12*).

To observe the chiral edge conduction on a single DW, the left-up-right-down domain structure is prepared with domain writing technique by MFM as shown in the upper left schematic drawings of Fig. 2B. Here, the Hall resistance at the left (right) side $R_{13}$ ($R_{24}$) is $\sim +h/e^2$ ($\sim -h/e^2$), corresponding to the magnetization direction. As for the longitudinal resistance, we note that $R_{12}$ and $R_{34}$ show completely different behavior from one another; $R_{12}$ takes a high resistance of $\sim 2h/e^2$, while $R_{34}$ is almost zero, which are in stark contrast to those in Fig. 2A, where $R_{13} = R_{24}$ and $R_{12} = R_{34}$. When a magnetic field is applied, the resistance values retrieve those of the single-domain state, confirming its magnetic domain origin. The values of $2h/e^2$ and $0h/e^2$ for the longitudinal resistance are explained by the following physical picture: At the DW of a magnetic TI in QAHE, two CESs are predicted to travel in the same direction as schematically shown in the black arrows in Fig. 2B due to the discontinuous change in Chern number between up and down domains (*2,3*). Here, the two parallel channels intermix and equilibrate with each other across the DW (*5,6*). Accordingly, the potentials at the downstream becomes equal so that $R_{34}$ becomes zero. In contrast, $R_{12}$ becomes twice the von Klitzing constant, $2h/e^2$, because only half of the electrons injected from the current contact 5 would be ejected to the other current contact 6. The reversal of the relationship between $R_{13}$ and $R_{24}$ ($R_{34}$ and $R_{12}$) in the right-up-left-down domain structure (Fig. 2C) is due to the chirality inversion of the CES. Therefore, these observations evidence the existence of the CES at the DW.

To further confirm the existence of the CES at the DW, we study the DW position dependence of the transport property as shown in Fig. 3C. As the tip scans over the device from left to right as shown in Fig. 3A, the magnetization direction is reversed little by little from down (blue) to up (red), and accordingly the DW changes its position; the DW position ($x$) is defined as the distance measured from the left contact 5 edge, as shown in Fig. 3B. Upon the nearly continuous drive of the DW, the Hall resistance $R_{13}$ ($R_{24}$) in Fig. 3C changes from ~ $-h/e^2$ to ~ $+h/e^2$ at the corresponding voltage contacts. Interestingly, $R_{12}$ takes a peak structure at ~ $2h/e^2$ only when the DW is in between the two contacts, consistent with the consideration from the chiral edge conduction along the DW. When the DW reaches the right-side end, the resistance values get back to those of the single domain state. The DW position dependence is also clearly demonstrated in the control experiments (fig. S4).

The CES width can be estimated from the resistance transition width in Fig. 3C. If the CES width were narrow enough, the Hall resistance $R_{13}$ would take a constant value when the DW position is away from the voltage contacts 1 and 3. However, this is not the case in the actual experiment. We attribute this behavior to the finite width of the CES at the DW; if the CES has some spatial distribution, it starts to affect the electrical potential and Hall resistance even when the DW is not on the voltage contacts. Here, we tentatively fit a part of the resistance value $R_{13}$ with an exponential function of the DW distance away from the voltage terminals 1 and 3: The fitting procedure gives the CES width to be approximately 5 μm, much larger than DW width itself judged from the MFM image in Fig. 1F. In the multi-domain state, the CES width is much larger than the typical domain size of ~ 200 nm (Fig. 1D), so that conduction occurs through the tunneling between the multiple CESs (*18-20*), leading to the continuous change of resistance as observed in Fig. 2A. Therefore, writing a large magnetic domain is crucial for the observation of the CES at the DW in the present device.

Finally, we show the proof-of-concept experiments of the CES-based electronic devices. We write various domain patterns by the MFM tip and measure the transport properties, which are summarized in Fig. 4A. In the case of the ideal chiral edge conduction, one can calculate the resistance using Landauer-Büttiker formalism (*21*) as shown in the horizontal solid bars. Because of the dissipation-less one-dimensional nature of the CES, the resistance value is determined just by the relative positions of the DWs, independent of the actual size of the device. Although there are some deviations due to the finite CES width and the Hall-bar size, the experimental values show good agreement with the ideal values. Importantly, in contrast to the case of the QHE (*1,4-6*), these multiconfiguration can be freely interchanged by the local magnetic field; namely these states are reconfigurable within a single device. Further demonstrations are displayed in Fig. 4, B and C, presenting the possible devices operation with two-terminal resistance. Remarkably, the behaviors are qualitatively different depending on the DW configuration: when the DWs connect the current contacts (Fig. 4B), the resistance decreases with increasing the number of DWs. In contrast, when the DWs connect the sample edges (Fig. 4C), the resistance increases with increasing the number of DWs. This is consistent with the Landauer-Büttiker formalism (*21*), because the ideal resistance is quantized to $R_{56} = h/(N+1)e^2$ in the former case and to $R_{56} = (N+1)h/e^2$ in the latter case, where $N$ is the number of DWs. This behavior means that the CES works as a conductive channel when it connects the two current contacts, while it rather works as an edge channel reflector between the CES at the sample edges when it connects the two sample edges. The qualitative difference originates from the chiral nature of the CES and is clearly distinct from conductive DWs driven by different physical origins (*22-25*) *e.g.* those found in ferroelectric insulators (*22,23*) or magnetic insulator with all-in-all-out magnetic order (*24,25*).

In conclusion, our observation clearly proved the existence of the CES on a magnetic DW of a magnetic TI through the domain writing by MFM on a mesoscopic scale. In addition, the proof-of-concept CES devices was demonstrated based on their unique chiral nature. The present discovery, combined with the recent spintronic developments based on the large spin-orbit torque due to the spin-momentum locking of the surface state of TI (*13,14*), would enable all-electrical control of the mobile DW and the CES, leading to the low power-consumption CES-based logic and memory devices (*22,26-28*) and quantum information processing (*3,29,30*) in the future.

**Acknowledgments:** We thank M. Kawamura for fruitful discussions. K. Y. is supported by the Japan Society for the Promotion of Science (JSPS) through a research fellowship for young scientists (No. 16J03476). This research was supported by the Japan Society for the Promotion of Science through the Funding Program for World-Leading Innovative R & D on Science and Technology (FIRST Program) on "Quantum Science on Strong Correlation" initiated by the Council for Science and Technology Policy and by JSPS Grant-in-Aid for Scientific Research(S) No. 24224009 and No. 24226002 and No. JP15H05853 from MEXT, Japan, CREST, JST (Grant No. JPMJCR16F1).

K. Y., M. M. and R. Y. grew the topological insulator heterostructures and performed device fabrication with the help of A. T. and K. S. T.. K. Y. and F. K. conducted the MFM measurement and analyzed the data. K. Y., A. T., M. K., F. K. and Y. T. jointly discussed the result and wrote the manuscript with contributions from all authors.

**Fig. 1**. **Magnetic domain writing by MFM in a magnetic topological insulator.** (**A**) Schematic illustration of CESs (black arrows) at the magnetic DW of a magnetic TI. $M > 0$ and $M < 0$ indicate upward (red) and downward (blue) spontaneous magnetization, respectively. (**B**) The surface band structures corresponding to the magnetic structures. When the magnetization points perpendicular to the film, exchange interaction opens up the mass gap. At the DW between up and down magnetic domains, two gapless CESs appear because of the discontinuous change of the Chern number from $C = +1$ to $-1$. (**C**) Schematic drawing of the Cr modulation-doped magnetic TI. BST and Cr-BST stand for $(Bi_{1-y}Sb_y)_2Te_3$ and $Cr_x(Bi_{1-y}Sb_y)_{2-x}Te_3$, respectively, where the nominal compositions are $x \sim 0.2$ and $y \sim 0.74$. (**D**) Magnetic domain structure of the naturally-formed multi-domain state around the magnetization reversal point at $B = 0.059$ T. The measurement is performed using MFM with a non-contact mode. The color scale shows the resonance frequency shift $df$, where the red and blue colors correspond to up and down domains, respectively. The color scale corresponds to 0.50 Hz span from red to blue. The scanned area is 1.5 μm × 1.5 μm, where the scale bar indicates 500 nm. (**E**) Schematic illustration of the domain writing procedure using MFM with a contact mode. The stray field from the MFM tip reverses the magnetization direction of the sample as shown in the enlarged circle (the red curves represent the stray field). (**F**) Magnetic domain structure after the domain writing procedure as shown in **e**. The magnetization direction is reversed only within the dotted line frame (0.75 μm × 0.75 μm area) scanned with a contact mode. Finite contrast in single domain regions originates from the spatial inhomogeneity of non-magnetic forces such as electrostatic forces.

**Fig. 2. Observation of chiral edge conduction on a magnetic domain wall.** (**A**) Magnetic field dependence of the Hall resistance $R_{13} = R_{24}$ and the longitudinal resistance $R_{12} = R_{34}$ at $T = 0.5$ K.

The upper illustration shows the corresponding magnetization configuration and the chirality of CES. Current is injected from contact 5 to contact 6 and voltage is measured at contacts 1-4. Note that the multi-domain state at the upper middle figure is just a schematic view, and the actual size of the domain (typically ~ 200 nm) is much smaller than the device size (several tens of µm). (**B**) Transport property of the left-up-right-down domain structure (upper left figure) and the subsequent magnetic field dependence. (**C**) Transport property of the left-down-right-up domain structure (upper left figure) and the subsequent magnetic field dependence.

**Fig. 3. Domain wall position dependence of the transport property.** (**A**) Schematic drawing of a measurement procedure for the DW position dependence of the transport property. By scanning the sample with the MFM tip under an external magnetic field of 0.015 T, the DW position is continuously moved from the left to right. (**B**) An optical micrograph of the device: modulation-doped magnetic TI (green), gold electrodes (blown). The size of the device is 24 µm × 48 µm, where the scale bar indicates 10 µm. (**C**) The DW position $x$ dependence of the transport property. The shaded area denotes the three typical cases of DW position relative to the voltage contacts as schematically displayed in the upper figures. The black line represents the fitting of $R_{13}$ with an exponential function $R_{13} = R_0 - re^{-(x-9\mu m)/\xi}$, where the fitting gives $R_0 = 23 \text{ k}\Omega$, $r = 16 \text{ k}\Omega$ and the CES width $\xi = 5.0 \ \mu m$.

**Fig. 4. Proof-of-concept experiments of the CES-based electronic devices.** (**A**) DW configuration dependence of the four-terminal resistance $R_{13}$ (light red), $R_{24}$ (dark red), $R_{12}$ (light blue), $R_{34}$ (dark blue) and the two-terminal resistance $R_{56}$ (light green) at $B = 0$ T. The resistance is shown in the unit of $h/e^2$. The ideal resistance values derived from the Landauer-Büttiker formalism are indicated by horizontal solid bars. Note that only the $R_{56}$ value in the last configurations takes a half integer value of 1/2. (**B**) Two-terminal *I-V* characteristics measured with various DW configurations, where DWs connect the current contacts (brown). The inset figures show the legends of the measurement configurations. The solid line and the dotted line almost overlap as expected from symmetry. (**C**) The same as (B) for the DWs connecting the sample edges.

**Supplementary Materials:**

Materials and Methods

Figs. S1 to S5

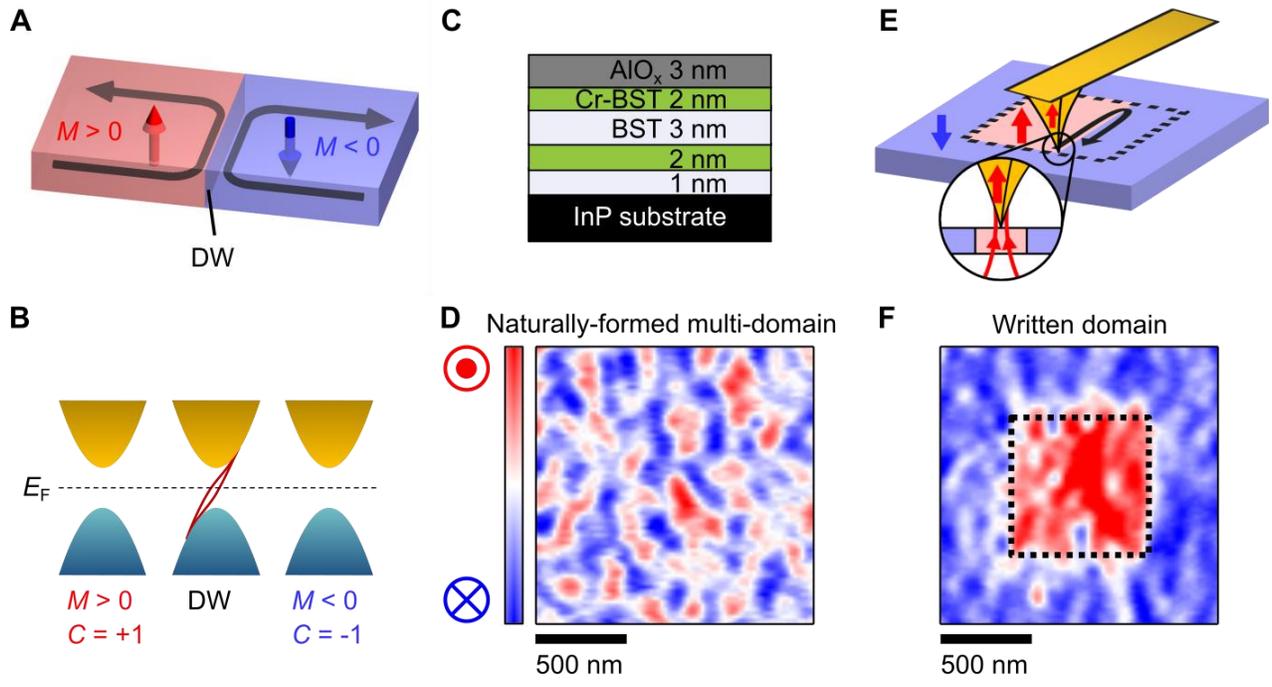

Fig. 1 K. Yasuda *et al.*,

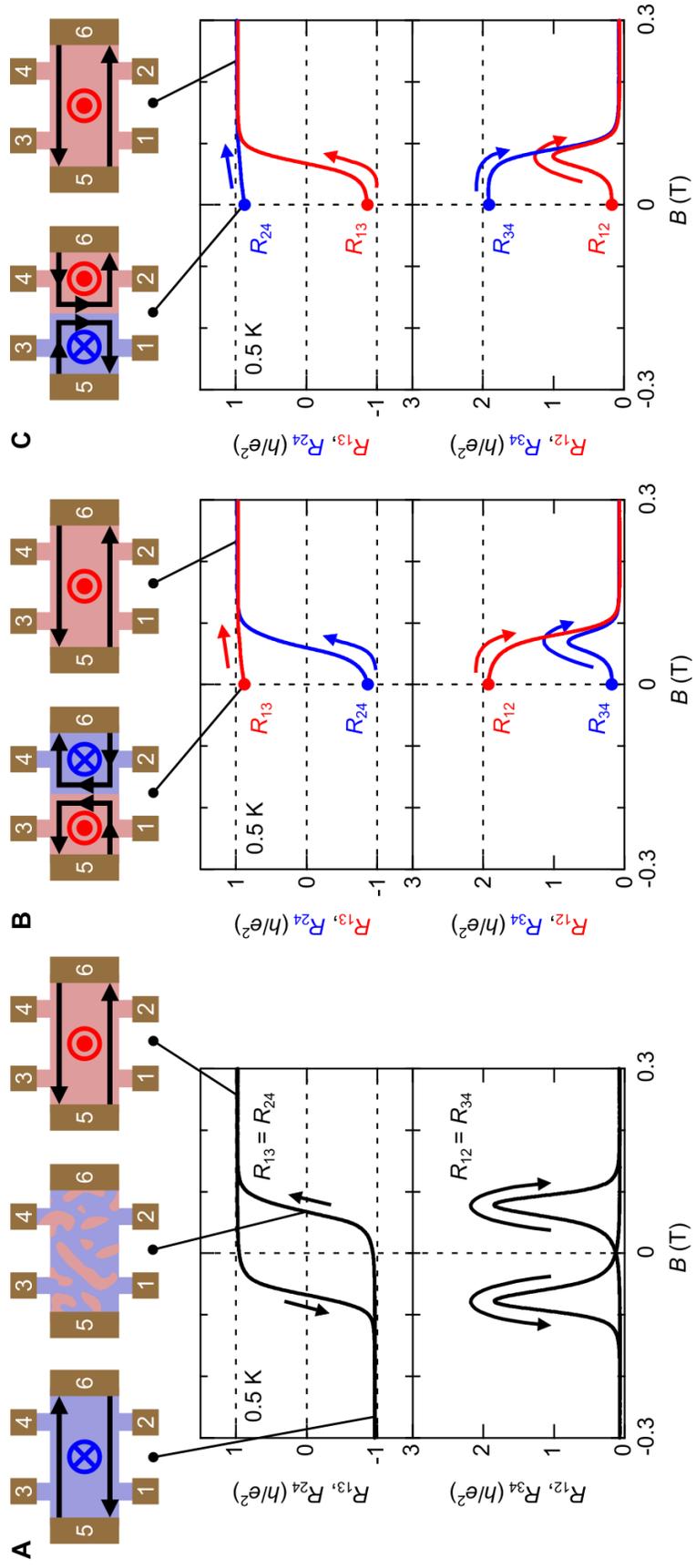

Fig. 2 K. Yasuda et al.,

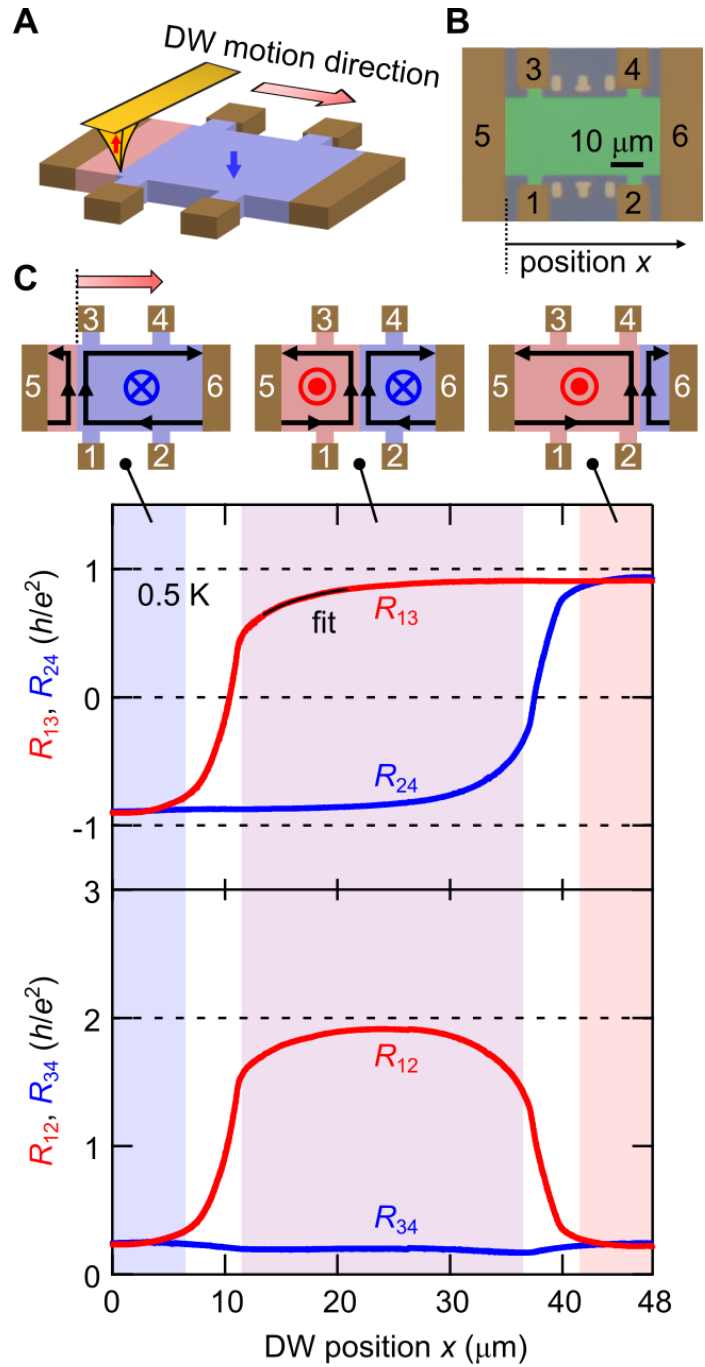

Fig. 3 K. Yasuda *et al.*,

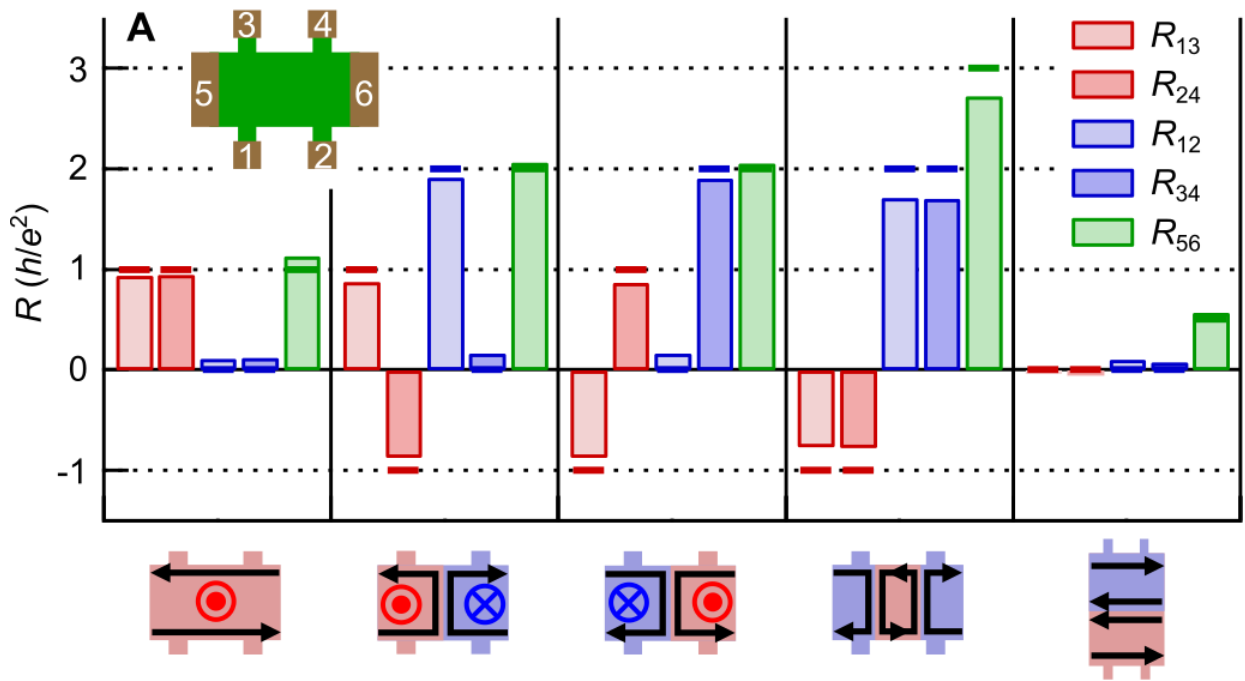
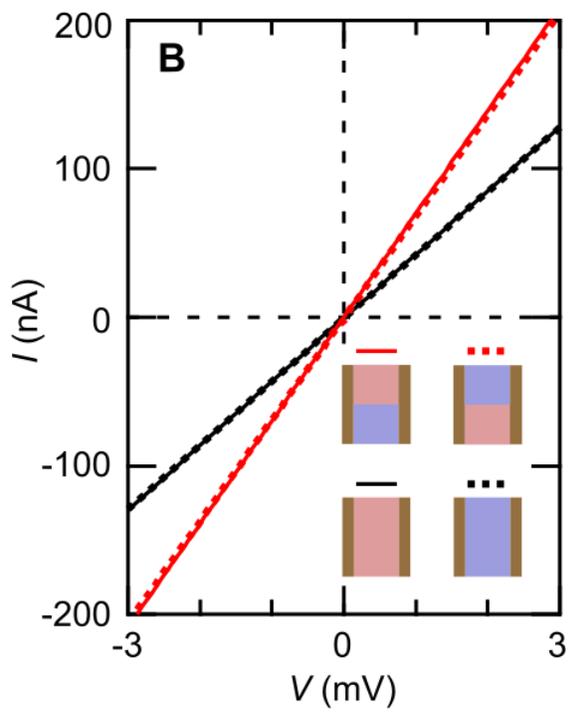
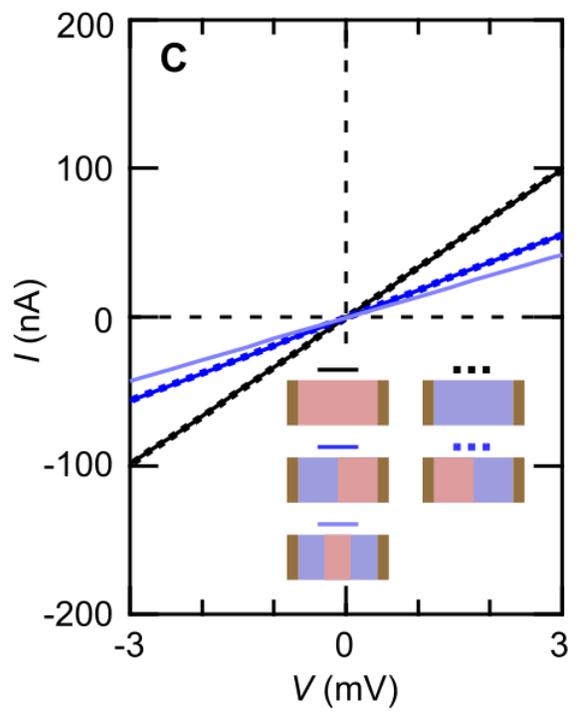

Fig. 4 K. Yasuda *et al*.,